\documentclass[epj,twocolumn,english]{webofc}

\usepackage[varg]{txfonts}   
\newcommand{\invisible}[1]{}

\def \etmis {\mbox{\ensuremath{E\kern-0.6em\slash_T}}}
\def \etmiss {\mbox{\ensuremath{E\kern-0.6em\slash_T}}}

\newcommand{\rien}{}

\newcommand{\INVISIBLE}[1]{}

\newcommand{\dzero}  {\ensuremath{\mathrm{D\O}}}

\newcommand{\sqrts}{\ensuremath{\rien{\sqrt{s}}}}

\newcommand{\gev}  {\ensuremath{\mathrm{ GeV}}}

\newcommand{\egev}{\gev}

\newcommand{\pbinv}{\ensuremath{\mathrm{ pb^{-1}}}}

\newcommand{\invfb}{\ensuremath{\mathrm{ fb^{-1}}}}

\def\eg{{\it e.g.}}

\newcommand{\pt}{{\ensuremath{ p_T}}}

\newcommand{\ipb}  {\pbinv}

\newcommand{\BEA}{\begin{eqnarray}}
\newcommand{\EEA}{\end{eqnarray}}

\def\beq  {\begin{equation}}
\def\eeq  {\end{equation}}

\renewcommand{\dzero}{D0}
\woctitle{LHCP 2013}

\begin{document}
\selectlanguage{english} 

\title{Tevatron Higgs results}

%
%

\author{Boris Tuchming\inst{1}
\fnsep\thanks{\email{tuchming@cea.fr
    }} for the CDF and \dzero\ Collaborations.
}

\institute{
CEA Saclay -  Irfu/SPP - France.
         }

\abstract{%

We present the combination of searches for the Standard Model Higgs boson,
using up to 10~\invfb\ of $p\bar p$ collisions at  $\sqrts$=1.96~TeV 
collected with the CDF and \dzero\ detectors at the Fermilab Tevatron collider.
The major contributing channels are optimized  for the main production modes, the associated production with a vector boson ($VH$, with $V=W,Z$),
the vector boson fusion, and the gluon-gluon fusion,
and  the different decay modes $H\to b\bar b$, $H\to\tau^+\tau^-$, $H\to W^+W^-$, and $H\to \gamma\gamma$.
A significant excess of events is observed in the mass range $115<M_H<140~\gev$. The local significance corresponds to 3.0 standard deviations at $m_H=125~\gev$, consistent
with the mass of the new particle observed at the LHC.
The observed signal strengths in all channels are consistent with the presence of a standard model Higgs boson of mass 125~\gev.
We also present prospects for spin/parity tests to be performed in the $VH \to V b\bar b$ channels.

}

\maketitle
\section{Introduction }

\label{intro}

Finding the standard model (SM) Higgs boson has been of the most topical goals of particle physics in the last decades.

Until 2000, direct
searches were conducted at the CERN $e^+e^-$ collider (LEP) that finally yielded the lower limit of
$M_H > 114.4~\egev$~\cite{bib:lephiggs}.
This exclusion limit and those reported hereafter are all defined at the 95\% C.L.
After the end of searches at LEP, precision electroweak tests, including the $W$-boson mass mesurement from LEP, and the $W$-boson
 mass and top-quark mass measurements from the Tevatron Run~I, 
contributed to constrain further the Higgs boson mass. In Summer 2002, the  constraint from the electroweak fit read $M_H<193~\egev$~\cite{ALEPH:2002aa}.
%
This was greatly improved thanks to the top mass and $W$ mass measurements using the first data of Tevatron Run~II,
so that in Winter 2007, the indirect constraint  was
$M_H<144~\egev$~\cite{Barberis:2007zz}
which narrowed down substantially the expected mass range for the Higgs boson.

With the beginning of Run~II of the Tevatron $p \bar p$ collider at $\sqrt s=1.96$~TeV, a new cycle of searches started in 2002, 
that ended with the final Tevatron shutdown in September 2011.
In 2008, the Tevatron Collaborations, CDF and \dzero,  presented the first post-LEP-era direct constraint on the Higgs, excluding
the mass $M_H$=170~\gev~\cite{Bernardi:2008ee}. This constraint
was extended over the years~\cite{bib:tevcomb-hww}, and for example at Summer 2011 conferences the  mass range from 156~\egev\ to 177~\egev\ was excluded~\cite{CDFandD0:2011aa}.

With the Large Hadron Collider (LHC) a new era started.
The 7 and 8~TeV $pp$ collision data from 2011-2012 allowed to explore a wide range of Higgs boson mass, and establish  more stringent
limits.
Upper (lower) limits of  131 (122)~\egev~\cite{bib:atlas-hdiscovery} and 128 (121.5)~\egev~\cite{bib:cms-hdiscovery} were obtained
by the ATLAS and CMS Collaborations, respectively.
But the perspective changed dramatically in 2012 with several announcements.
On July 4th 2012, the ATLAS and CMS Collaborations
reported  excesses above  
background expectations at the five standard deviation (s.d.) level,
consistent with the observation of a Higgs boson of $M_H \approx 125$~\egev~\cite{bib:atlas-hdiscovery,bib:cms-hdiscovery}.
In the same week, the CDF and \dzero\ Collaborations  reported excesses
above background expectations in the $H\rightarrow b\bar b$
search channels~\cite{bib:CDFhbb,bib:D0hbb}. 
Combining CDF and \dzero\  yields an excess at the three s.d.\ level,
consistent with the production of a Higgs boson of mass $M_H \approx 125$~\egev~\cite{bib:TeVhbb}.
With the discovery of the new particle, a measurement time started.

This proceedings discusses the final combined search results from the Tevatron collaborations,  which represents the culmination of more than ten years of data analysis. 
For most of the channels, the full Run~II dataset is used which corresponds to $\sim 10$~\invfb\ 
of $p\bar p$ collisions per experiment after data quality requirements.
The results are also interpreted to measure properties of the newly discovered particles: production rate in different modes and measurement of couplings to fermions and bosons.
Most of these results have been recently submitted and accepted for publication (see Refs.~\cite{{cdfprd},{d0prd},{CDFDOcombo}} and references therein).
The \dzero\ internal combined results are discussed elesewhere in this proceedings~\cite{bib:Emilien}.

\section{Higgs boson production and decay channels}

In the SM, the production cross-sections and the branching fractions as a function of the  Higgs mass are well known. Over the mass range $90<M_H<200$~\egev, the dominant production process is the gluon-gluon fusion $gg\rightarrow H$ (950~fb at 125~\gev), followed by  the associated production with a weak vector boson $p\bar p\to WH,\ p\bar p\to ZH$ (130 and 79~fb at 125~\gev)
and the weak vector boson fusion $p\bar p\to q q' H$ (67~fb at 125~\gev).
The main decay modes for $M_H=125$~\gev\  are $H\to b\bar b$  (58\%) and $H\to W^+W^-$ (22\%).
The most sensitive signatures are:
\begin{itemize}
\item   one lepton + \etmis\ + two $b$-jets   ($WH\to \ell\nu b \bar b$)
\item   no lepton + \etmis\ + two $b$-jets   ($ZH\to \nu \bar\nu b \bar b$)
\item   two leptons  + two $b$-jets   ($ZH\to \ell^+\ell^- b\bar b$)
\item  two leptons + \etmis\  ($H\to W^+W^- \to  \ell^+\nu\ell^-\bar\nu$) 
\end{itemize}
Thus, the search for the SM Higgs at Tevatron mainly relies on
$b$-tagging efficiency, good dijet mass resolution, high-\pt\ lepton acceptance,
good modeling of the \etmis, and good modeling of the $V$+jet background.
The Tevatron sensitivity to $VH\to V b\bar b$ is complementary to the LHC main discovery channels ($H\to\gamma\gamma, H\to ZZ$), which should help unravel the nature of the new particle.

The main sensitivity is given by the four channels presented above, but many other signatures are also considered
to bring additional sensitivity and test the agreement with the SM expectations.
For examples, Tevatron experiments have also looked for diphoton events ($H\to\gamma\gamma$), associated production with top-quark pairs ($t\bar t H$),
lepton + \etmis\ + dijet signature (from $H\to WW$),
 trilepton signatures (\eg\ from $WH\to WWW$), same charge dilepton signatures (\eg\ from $WH\to WWW$),
quadrilepton signatures (\eg\ from $ZH\to \ell^+\ell^- WW\to \ell^+\ell^-\ell^+\ell^-\nu\bar\nu$),
and tau-based signatures (\eg\ from $WH\to \ell\nu\tau^+\tau^-$ or $WH\to q\bar q \tau^+\tau^- $).

\section{Analysis strategy}
Over the course of Run~II, both collaborations have followed the same strategy to
optimize the analyses and improve
their sensitivity faster than expected from just accumulating more and more data.
\begin{itemize}
\item  Acceptance is maximized
by lowering kinematic requirements on leptons, by including different lepton reconstruction categories,
by accepting events from all possible triggers, and by optimizing $b$-jet tagging with more and more sophisticated multivariate techniques (MVA).
\item 
MVA techniques are widely used in all channels as they provide typically 25\% more sensitivity
than just using  single kinematic discriminant such as the dijet mass for the $VH\to Vb\bar b$ channels.
The improved sensitivity obtained thanks to MVA can be assessed by eye in Fig.~\ref{fig:d0_vvbb_mva},
which compares the most discriminant variables (dijet mass) and the MVA in the  \dzero\ $\etmis+b\bar b$ search channel~\cite{{dzZHv2}}.
\begin{figure}[!htb]
\centering
\includegraphics[width=0.45\textwidth,clip]{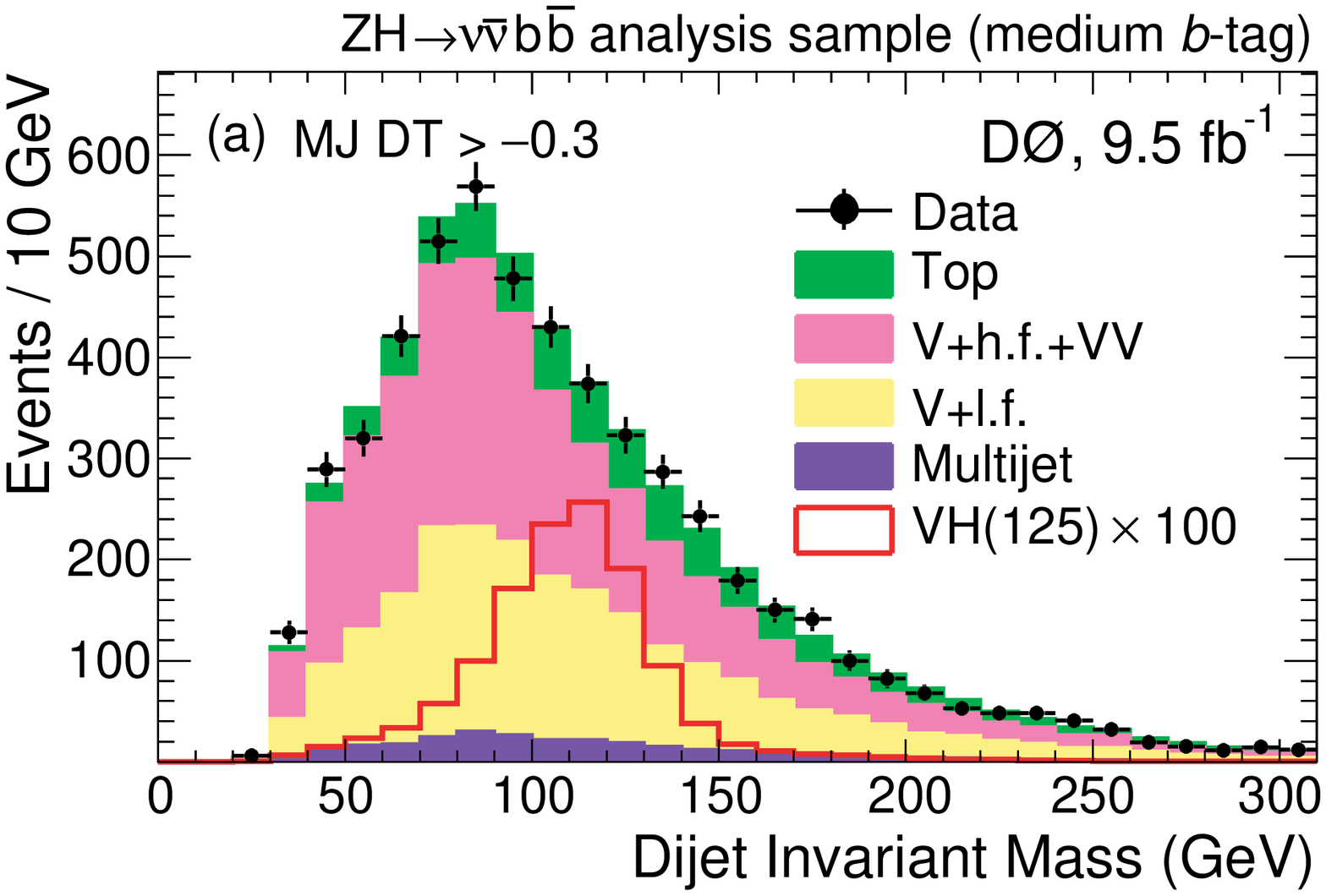}
\includegraphics[width=0.45\textwidth,clip]{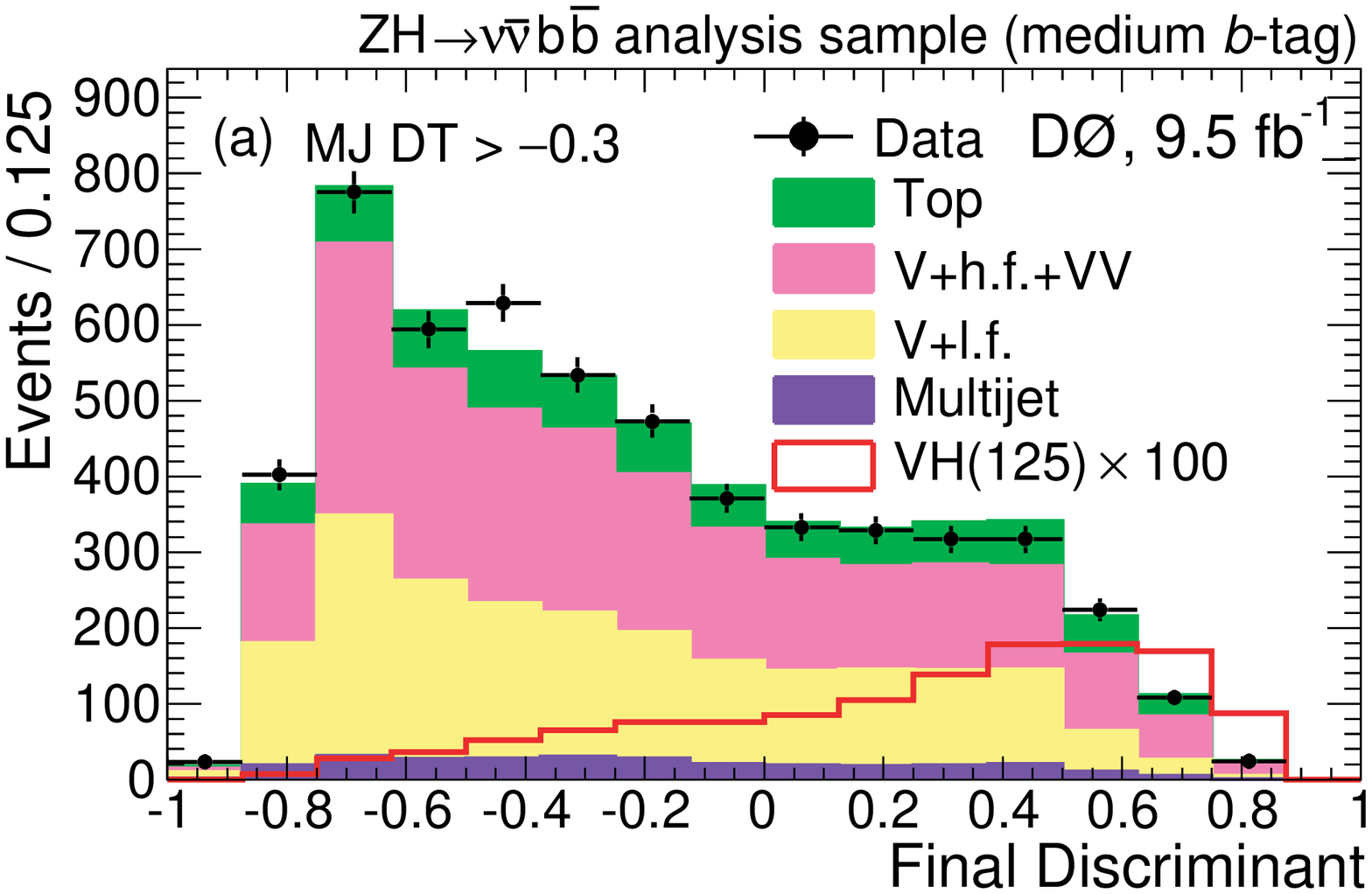}
\caption{Top, distribution of the dijet invariant mass of candidate events in the \dzero\ $\etmis+b\bar b$ search channel.
Bottom, distributions of the final MVA at the same selection stages.}
\label{fig:d0_vvbb_mva}       
\end{figure}

Each MVA combines into a single discriminant many variables,
which include for example variables describing the
event topology, the lepton
and jet kinematics, the quality of leptons, and the relation
between leptons/jets and \etmis. All channels at Tevatron employ at least one MVA, optimized for each different
Higgs boson mass hypothesis.
For most of the analyses
several MVAs trained specifically against different backgrounds bring additional sensitivity.

The MVA techniques are also employed for object identification ($b$-jets, leptons, photons) and for energy correction to $b$-jets. For example the usage of the CDF
{\em HOBIT}~\cite{bib:hobit} b-tagging algorithm
in the final published $VH\to V b\bar b$
analyses provides an enhancement of $\sim20\%$ in $b$-tagging efficiency per jet.

\item
Another way of achieving better sensitivity to signal,
consists in splitting the search channels into subchannels according to
jet multiplicity, $b$-tagging content, lepton flavor or lepton quality. Dedicated MVA are also trained to split analyses into subchannels enhanced or enriched in specific backgrounds. 
Using subchannels with different signal-over-background ratio ($s/b$) maximizes discriminating power, allows sensitivity to different signal production modes, and give more handles and lever-arm to control backgrounds and systematic uncertainties.
As an example, Fig.~\ref{figc_cdf_Metbb_subchannels}  shows the final MVA output for the three $b$-tagging categories of the CDF $\etmis+b\bar b$ analysis~\cite{cdfmetbb}.
\begin{figure}[!htb]
\centering
\includegraphics[width=0.45\textwidth,clip]{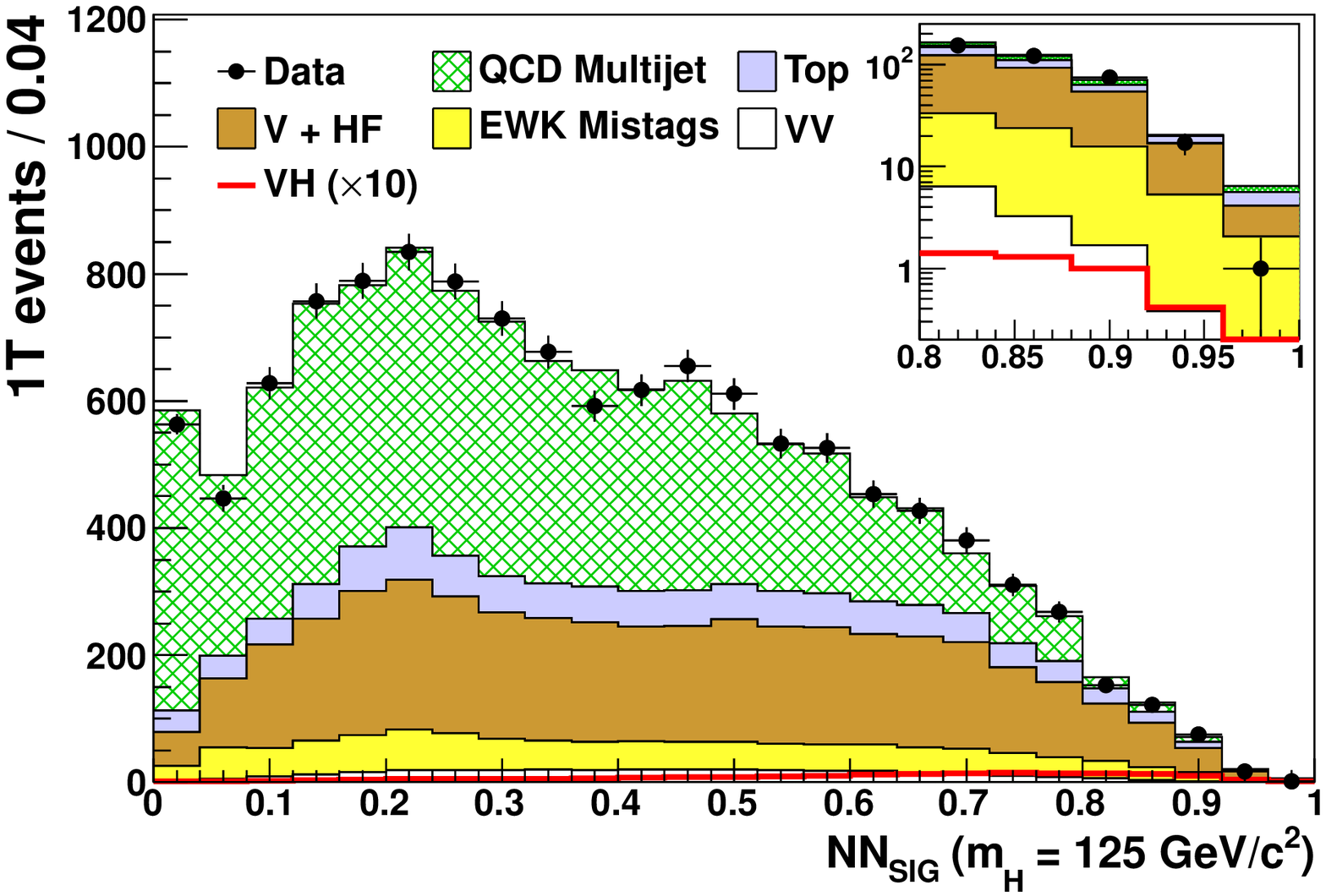}
\includegraphics[width=0.45\textwidth,clip]{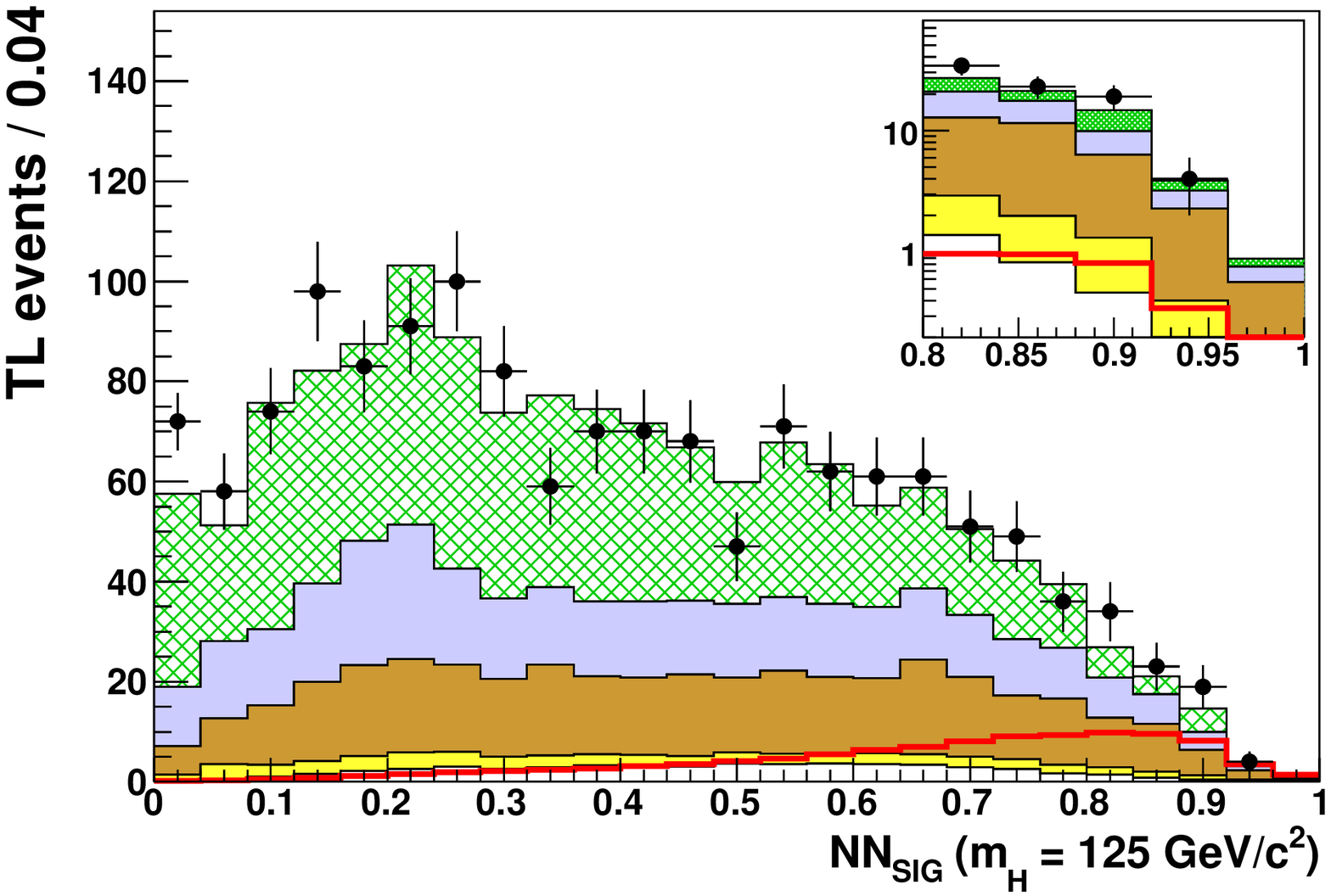}
\includegraphics[width=0.45\textwidth,clip]{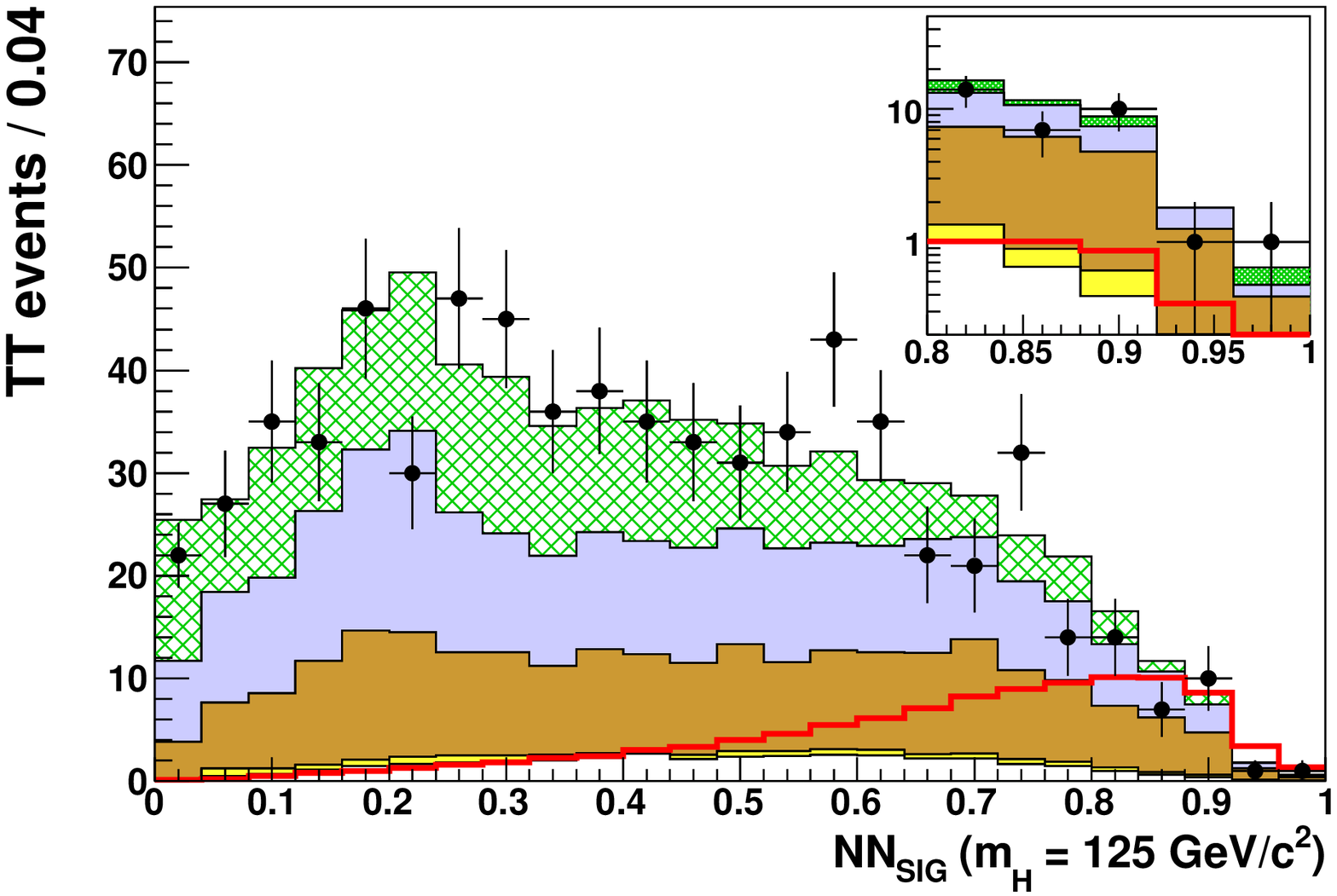}
\caption{Final MVA discriminant in CDF $\etmis+b\bar b$ analysis for the three
subchannels defined according to the $b$-tagging of jets:
top, only one jet is tightly tagged (1T);
middle,  one
jet is tightly tagged, and the other loosely tagged (TL);
and bottom, both jets are tightly tagged (TT).
 }
\label{figc_cdf_Metbb_subchannels}       
\end{figure}

\item
The data are employed as much as possible to control backgrounds and validate search methods. Instrumental backgrounds, such as jets faking leptons, photons faking electrons, charge mismeasurements, and tail of \etmis\ resolution are measured in dedicated control samples. Background enriched samples are also employed to check modeling of specific background processes.
Finally, the same analysis techniques, namely the same kind of MVA, the same subchannels and the same treatment of systematic uncertainties are employed to measure production rates of known standard model candles such $p\bar p\to W^+W^-\to \ell^+\nu\ell^-\bar\nu$, or $VZ \to V b \bar b$.
At \dzero, the measured  cross section $\sigma(W^+W^-)=11.6\pm0.7~\ipb$ is in agreement with the
NNLO prediction of $11.3\pm0.7~\ipb$.
The combined CDF+\dzero\ measured cross section $\sigma(WW+WZ)=3.0\pm 0.6\ \rm{stat} \pm 0.7\ \rm{syst}~\ipb$ is in agreement with SM prediction of $4.4\pm0.3~\ipb$.
Figure~\ref{fig:tev_wz_zz_mass} shows the background subtracted dijet mass distribution in highest $b$-tagging categories used in this later measurement.
\begin{figure}[!htb]
\centering
\includegraphics[width=0.45\textwidth,clip]{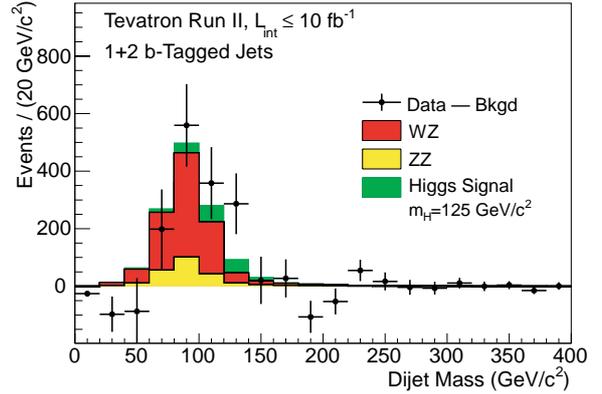}
\caption{Background subtracted dijet invariant mass distribution of the $VZ$ analyses in
the high $b$-tag purity categories.}
\label{fig:tev_wz_zz_mass}
\end{figure}
\end{itemize}

\section{Result of the search}
\begin{figure}[!htb]
\centering
\includegraphics[width=0.40\textwidth,clip]{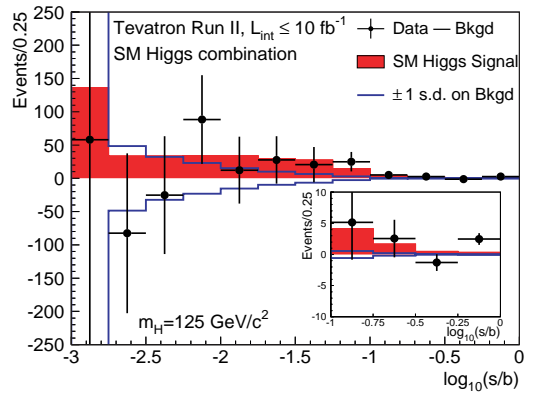}
\caption{Background subtracted distribution of events using
the final discriminant of all channels sorted in bins of $s/b$, for $M_H=125~\gev$.
}
\label{fig:tev_mva}
\end{figure}

\begin{figure}[!htb]
\centering
\includegraphics[width=0.45\textwidth,clip]{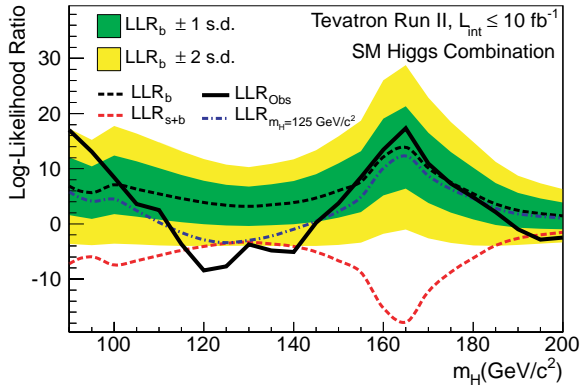}
\caption{LLR computed as a function of Higgs boson mass hypothesis.
}
\label{fig:llr}
\end{figure}

The combination of all search channels  for $M_H=125~\gev$ can be  visualized in Fig.~\ref{fig:tev_mva},
where the background subtracted distribution of the final discriminant for all channels are sorted as a function of $s/b$ and then added.
An excess of events in the highest $s/b$ bins is observed.
Figure~\ref{fig:llr} shows
the log-likelihood ratio (LLR) testing the signal-plus-background over the background-only hypothesis and computed for different test masses.

\begin{figure}[!htb]
\centering
\includegraphics[width=0.45\textwidth,clip]{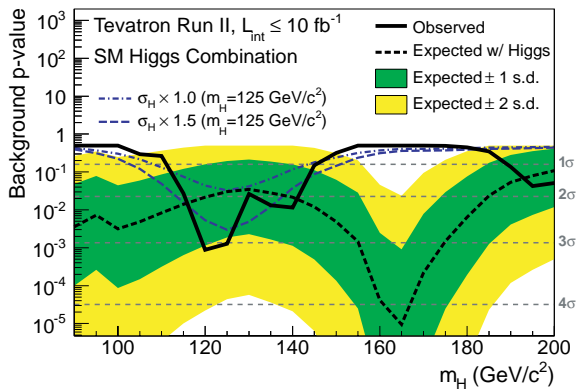}
\caption{Background p-value as a function of the Higgs boson mass hypothesis.}
\label{fig:tev_clb}
\end{figure}

A significant signal-like excess in the mass range between 115 and 140~GeV is observed.
The background  p-value of that excess corresponds to 3.0 s.d.\ for $M_H=125~\gev$, as shown in Fig.~\ref{fig:tev_clb}.
That excess arises from both CDF (2.0 s.d.) and \dzero\ (1.7 s.d) data, as expected from the presence of a standard model Higgs boson of mass 125~\gev.

\begin{figure}[!htb]
\centering
\includegraphics[width=0.45\textwidth,clip]{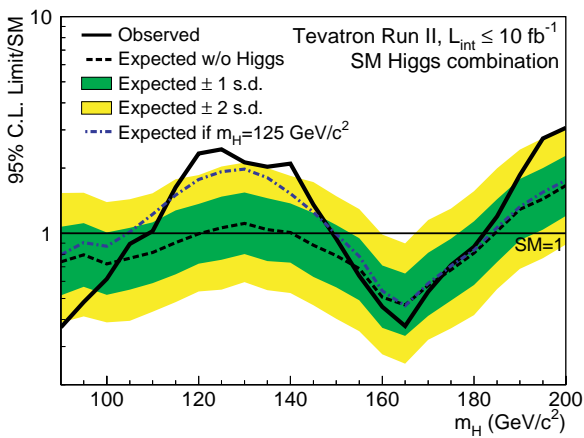}
\caption{Limits on the SM Higgs boson production as a function of the mass hypothesis.}
\label{fig:tevlimit}
\end{figure}

The results are interpreted in terms of limits on the Higgs boson production measured in units of the expected SM production.
They are shown in Fig.~\ref{fig:tevlimit} as a function of the Higgs boson mass.
The combined CDF and \dzero\ results have almost reached the exclusion sensitivity for the full range $[90,185]~\gev$. More precisely the expected excluded regions at 95\% C.L.\ are   $[90,120]~\gev$ and  $[140,184]~\gev$.
Because of the excess observed in the low mass region, the actual observed exclusion ranges are smaller: $[90,109]~\gev$ and  $[149,182]~\gev$.
For $M_h=125~\gev$, the observed (expected) production limits are $2.44\times$SM  ($1.06\times$SM).

\section{Searches beyond the standard model}

The  different SM channels can be used to search for physics beyond the SM.

The Tevatron Collaborations interpret their results in fermiophobic framework in which
the Higgs couplings to fermions are heavily suppressed,
thus suppressing the gluon-gluon fusion process.
In this search the main modes are the $H\to\gamma\gamma$ and $H\to WW$ search channels,
as both branching ratios are enhanced. 
The combined observed (expected) exclusion resulting from these channels is
$M_H<116~\gev$ ($M_H<135~\gev$).

\begin{figure}[!htb]
\centering
\includegraphics[width=0.45\textwidth,clip]{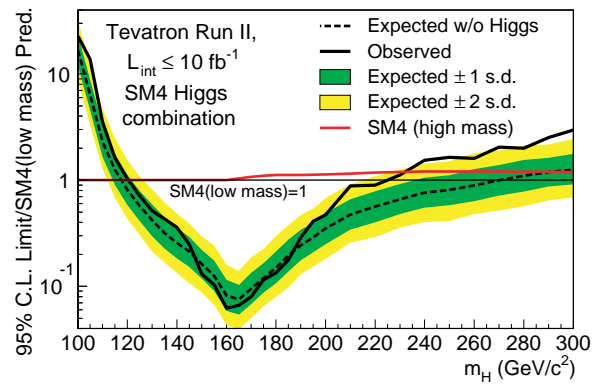}
\caption{Limits on the Higgs boson production in the context of a fourth generation of fermions as a function of the Higgs boson mass.}
\label{fig;fourth_gen_lmit}
\end{figure}

Another interpretation is performed within the context of a fourth generation of fermions.
In this framework the existence of heavy colored quarks
enhances the gluon-gluon fusion by approximately a factor of nine, thus the search
is performed in the $gg\to H \to WW$ channels only.
The absence of significant excess of data events allows to exclude
the mass range of $121<M_H<225~\gev$,
while the expected exclusion range is 
$118<M_H<270~\gev$, as shown in Fig.~\ref{fig;fourth_gen_lmit}.

\section{Higgs boson studies}
\subsection{Measurement of production rates}
\begin{figure}[!hb]
\centering
\includegraphics[width=0.45\textwidth,clip]{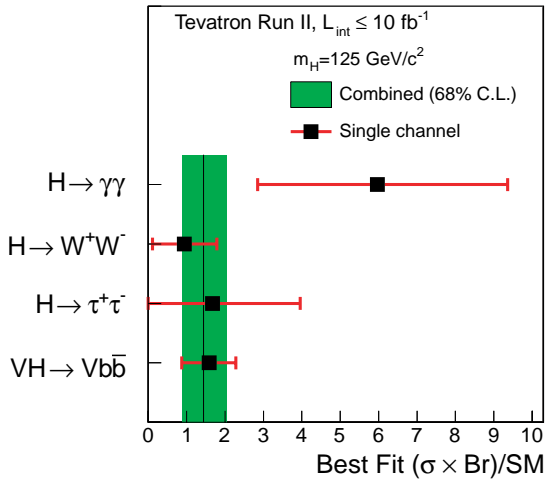}
\caption{Best fit to the data of the Higgs boson production in different channels, assuming  $M_H=125~\gev$.}
\label{fig:SM_yield}
\end{figure}

The SM search channels  can be separately combined to measure
the yield in the different modes: $H\to b\bar b$, $H\to\tau^+\tau^-$, $H\to W^+W^-$, and $H\to \gamma\gamma$. The best fits to the data are summarized in Table~\ref{tab:SM_yield} and displayed in Fig.~\ref{fig:SM_yield}. The overall production rate of
$1.44^{+0.59}_{-0.56}\times$SM  is obtained, compatible with the SM Higgs boson of mass 125~\gev. 
The modes with sizable signal-like excesses relative to the background-only hypothesis are  $VH\rightarrow Vb{\bar{b}}$  and $H\rightarrow W^+W^-$, as expected from the SM Higgs boson.

\begin{table}[!b]
\begin{tabular}{lcc|c}
   &  CDF~\cite{cdfprd}  & \dzero~\cite{d0prd}  &  CDF+\dzero~\cite{CDFDOcombo}\\\hline\hline
\rule[-2mm]{0mm}{7mm}
$R_{\rm{fit}}$(SM)                       &  $1.54^{+0.77}_{-0.73}$        &   $1.40^{+0.92}_{-0.88}$     & $1.44^{+0.59}_{-0.56}$ \\ \hline
\rule[-2mm]{0mm}{7mm}
$R_{\rm{fit}}$($H\rightarrow W^+W^-$)      & $0.00^{+1.78}_{-0.00}$      & $1.90^{+1.63}_{-1.52}$        & $0.94^{+0.85}_{-0.83}$  \\
\rule[-2mm]{0mm}{7mm}
$R_{\rm{fit}}$($VH\rightarrow Vb{\bar{b}}$) &$1.72^{+0.92}_{-0.87}$         & $1.23^{+1.24}_{-1.17}$      & $1.59^{+0.69}_{-0.72}$ \\
\rule[-2mm]{0mm}{7mm}
$R_{\rm{fit}}$($H\rightarrow \gamma\gamma$)&$7.81^{+4.61}_{-4.42}$       &  $4.20^{+4.60}_{-4.20}$     & $5.97^{+3.39}_{-3.12}$ \\ 
\rule[-2mm]{0mm}{7mm}
$R_{\rm{fit}}$($H\rightarrow \tau^+\tau^-$)&$0.00^{+8.44}_{-0.00}$        &         $3.96^{+4.11}_{-3.38}$ & $1.68^{+2.28}_{-1.68}$  \\
\rule[-2mm]{0mm}{7mm}
$R_{\rm{fit}}$($t\bar tH\rightarrow t\bar t b\bar b$)&$9.49^{+6.60}_{-6.28}$        &  --                       & --  \\
\hline\hline
\end{tabular}
\caption{
\label{tab:SM_yield}
Best fit to the data of the Higgs boson production (in unit of the SM Higgs boson production), assuming $M_H=125~\gev$,
for the different channels and their combination.
}
\end{table}

\subsection{Measurement of couplings to fermions and bosons}
We assume a SM-like Higgs particle of 125~\gev,
with no additional particle in loops and no invisible decays.
The SM couplings  to fermions and vector bosons are scaled by three numbers,
$\kappa_f$,
$\kappa_W$, and $\kappa_Z$. For example the $WH\to Wb\bar b$ yield is then scaled
by $\frac{\kappa_f^2 \kappa_W^2} {\kappa_H^2}$,
where the terms in the numerator correspond the scaling of the incoming and outgoing partial width, while the denominator is a global scaling factors for the total Higgs boson width.

A fit to the data is performed by separating and scaling properly the contributions from the different production and decay modes.
Note that in this procedure, only a few modes  exhibit some dependence on the relative sign between the coupling scale factors,
due to interferences  between diagrams.
The most important effect arises from the interference between $W$ loops and top-quark loops for the $ H\to\gamma\gamma$ partial width:
$\Gamma(H\to\gamma\gamma)= \Gamma(H\to\gamma\gamma)_{SM}\times (1.28\kappa_W -0.28\kappa_f)^2$. 

\begin{figure}[!htb]
\centering
\includegraphics[width=0.4\textwidth,clip,trim=0cm 1.6cm 0cm 1.6cm]{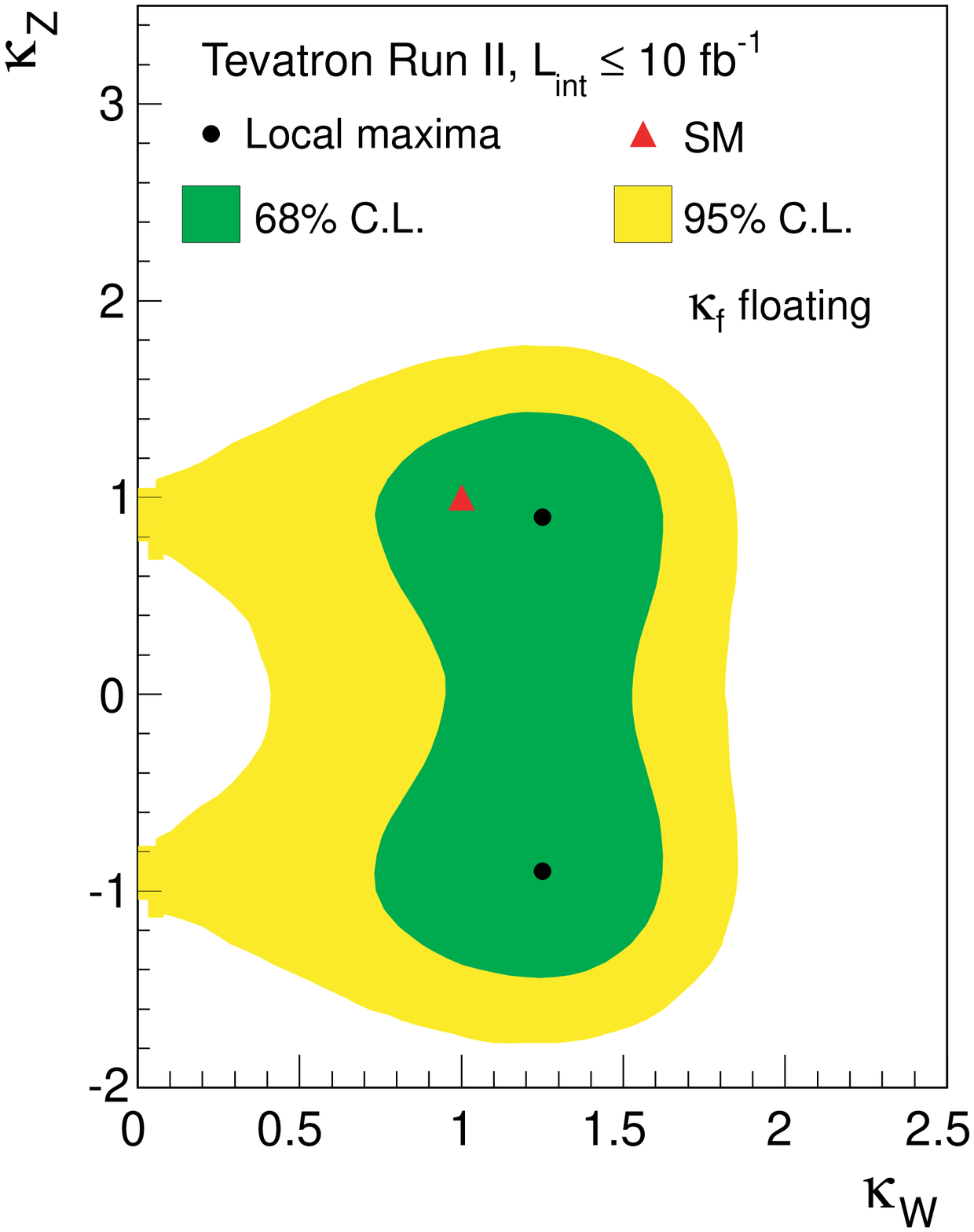}
\includegraphics[width=0.4\textwidth,clip,trim=0cm 1.6cm 0cm 1.6cm]{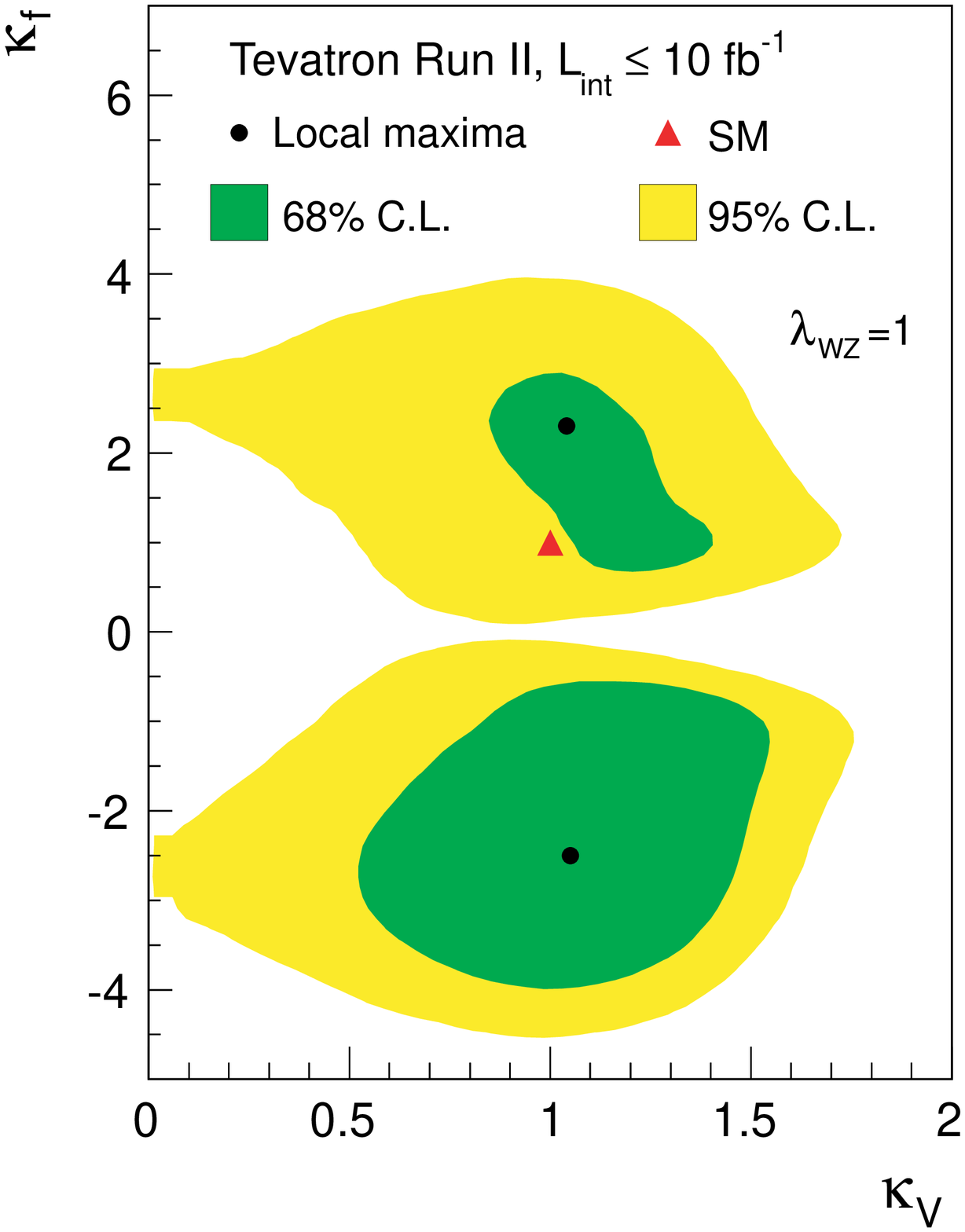}
\vspace{.5cm}
\caption{Posterior density probability for the coupling scale factors,
$ \kappa_f$,
$\kappa_W$, and $\kappa_Z$,  obtained under different assumptions.}
\label{fig:tev_coupling}       
\end{figure}

The results of the  fit are shown in Fig.~\ref{fig:tev_coupling}: 
when $\kappa_f$ is let floating (with flat prior) the best fit region is around  ($\kappa_W,\kappa_Z)=(1.25,\pm 0.90)$;
assuming custodial symmetry ($\kappa_W=\kappa_Z=\kappa_V$), the two preferred regions are around
($\kappa_V,\kappa_f)=(1.05,-2.40)$ and ($\kappa_V,\kappa_f)=(1.05,2.30)$.

These 2-dimensions results can be turned into one dimension constraints:
assuming $\kappa_W=\kappa_f=1$, the best-fit value is $\kappa_Z=\pm1.05^{+0.45}_{-0.55}$;
assuming $\kappa_Z=\kappa_f=1$,  the best-fit 68\%\ confidence intervals
are defined by  $\kappa_W=- 1.27^{+0.46}_{-0.29}$ and      $1.04<\kappa_W <1.51$;
assuming $\kappa_W=\kappa_Z=1$, the best-fit value
is $\kappa_f=-2.64^{+1.59}_{-1.30}$;
and by letting $k_f$ floating with a flat prior, 
the custodial symmetry is tested and the best fit value for the ratio
 $\lambda_{WZ}=\frac{\kappa_W}{\kappa_Z}$ reads  $\lambda_{WZ}=1.24^{+2.34}_{-0.42}$.
All these results are in agreement with the SM expectations within their uncertainties.

\subsection{Spin and  parity tests}
In general spin/parity of a particle affects angular distributions of its decay products,
but also cross-section behavior near production threshold. This later 
property can be exploited
at Tevatron in the $VH\to V b\bar b$ search modes.
The spectra of the effective center-of mass energy, $\sqrt{\hat s}$,
of $VH\to V b\bar b$ events are expected to be quite different under different
spin and parity hypothesis ($0^-$, $0^+$, or $2^+$) for $H$~\cite{bib:Elis}.
This  can be exploited by using as  main discriminant observable the 
overall mass (or transverse mass for final state with neutrinos) of the candidate events.
The \dzero\ analysis is discussed
elsewhere in this proceedings~\cite{bib:Boris2}. No measurement has been performed yet, but each
Tevatron Collaboration is expected to release results for the forthcoming conferences.

\section{Conclusion}

After ten years of excellent performance for
the Tevatron collider and the CDF and \dzero\ experiments,
both Tevatron collaborations  combine their final results on the SM Higgs boson searches.
They almost achieve exclusion sensitivity over the full range $[90-185]~\gev$,
and exclude at 95\%\ C.L. the range of mass $90<M_H<109~\gev$ and $149<M_H<182~\gev$.
Interpretation of results beyond standard model yields the limits
$M_H<116~\gev$ for a fermiophobic Higgs, and  $121<M_H<225~\gev$ in the context of a fourth generation of fermions.

In the search for the SM Higgs boson, both CDF and \dzero\ observe an excess of signal-like events
in the low mass range $115<M_H<140~\gev$,
compatible with the experimental resolution.
Its combined significance is 3.0 s.d. for $M_H=125~\gev$ and it arises mainly
from the $H\to b \bar b$ and $H\to W^+W^-$ channels, as expected
from the SM Higgs boson.
The measured  production rate of $1.44^{+0.59}_{-0.56}\times$SM and
the measured couplings are compatible with a Higgs boson of 125~\gev.
The experiments have also good prospects to probe the spin/parity
of the Higgs-like particle of 125~\gev\  in the $VH\to Vb\bar b$ modes.

\section{Acknowledgments}
%
We thank the Fermilab staff and technical staffs of the participating institutions for their vital contributions.  
We acknowledge support from the
DOE and NSF (USA),
ARC (Australia),
CNPq, FAPERJ, FAPESP and FUNDUNESP (Brazil),
NSERC (Canada),
NSC, CAS and CNSF (China),
Colciencias (Colombia),
MSMT and GACR (Czech Republic),
the Academy of Finland,
CEA and CNRS/IN2P3 (France),
BMBF and DFG (Germany),
DAE and DST (India),
SFI (Ireland),
INFN (Italy),
MEXT (Japan),
the Korean World Class University Program and NRF (Korea),
CONACyT (Mexico),
FOM (Netherlands),
MON, NRC KI and RFBR (Russia),
the Slovak R\&D Agency, 
the Ministerio de Ciencia e Innovaci\'{o}n, and Programa Consolider-Ingenio 2010 (Spain),
The Swedish Research Council (Sweden),
SNSF (Switzerland),
STFC and the Royal Society (United Kingdom),
the A.P. Sloan Foundation (USA), 
and the EU community Marie Curie Fellowship contract 302103.
%

%
%
%

\end{document}